\documentclass[journal=jpccck,manuscript=article,letterpaper,top=2cm,bottom=2cm,left=3cm,right=3cm,marginparwidth=1.75cm]{achemso}

\usepackage[english]{babel}
\usepackage{amsmath,amssymb}
\usepackage[export]{adjustbox}
\usepackage{graphicx}
\usepackage[colorlinks=true, allcolors=blue]{hyperref}
\usepackage{xcolor}
\usepackage[percent]{overpic}
\usepackage{float}
\usepackage{subcaption}
\usepackage[numbers,super,sort&compress]{natbib}

\title{Radial asymmetry in quadrupole mass filters: stability, multipole fields and resolution enhancement}

\author{Sukanya Jana}
\affiliation{School of Chemical Sciences, Indian Association for the Cultivation of Science, Kolkata-700032, India}

\author{Snigdha Bose}
\affiliation{Postgraduate and Research Department of Physics, St Xavier's College (Autonomous), Kolkata-700016, India}

\author{Sayel Chakraborty}
\affiliation{School of Chemical Sciences, Indian Association for the Cultivation of Science, Kolkata-700032, India}
\altaffiliation{Current address: Indian Institute of Technology Palakkad, Palakkad-678623, India}

\author{Pintu Mandal}
\affiliation{Department of Physics, St. Paul’s Cathedral Mission College, Kolkata-700009, India}
\email{pintuphys@gmail.com}

\author{Nabanita Deb}
\affiliation{School of Chemical Sciences, Indian Association for the Cultivation of Science, Kolkata-700032, India}
\email{nabanita.deb@iacs.res.in}

\begin{document}

\begin{abstract}

This study examines the effects of radial asymmetry in a linear quadrupole mass filter with circular rods, introduced either by a change in electrode radii or by displacement of a diametrically opposite pair. A radial potential model is developed to account for the resulting geometric deviations, enabling analysis of the first stability region specific to the quadrupole component. The transmission characteristics of such asymmetric configurations are systematically examined, revealing a linear shift in the transmission peak along the $q$-axis as a function of the asymmetry parameter. This behavior is interpreted through modifications observed in the stability diagram. Furthermore, increased asymmetry is shown to broaden the transmission contours and generally reduce resolution. Notably, a `magic' asymmetry parameter of $-0.02$ corresponding to electrode displacement, yields enhanced resolution compared to the symmetric case for a fixed rod-to-field radius ratio. An empirical relationship is established between the resolution and the combined contribution of the coefficients of octupole and dodecapole potential components, highlighting the critical role of higher-order field effects in performance optimization. 
\end{abstract}

\textbf{keywords}: Quadrupole mass filter, radial asymmetry, stability diagram, multipole fields, transmission characteristics, resolution

\section{Introduction} \label{sec1}

The quadrupole mass filter (QMF), introduced by Paul and Steinwedel in the 1950s~\cite{paul1953neues}, has become a mainstay in mass spectrometry due to its simplicity, robustness, and versatility in both stand-alone and hybrid instrument platforms \cite{paul1990electromagnetic,ghosh1995ion,march2005quadrupole,dawson2013quadrupole}. Over the decades, it has been the subject of extensive theoretical, experimental, and computational investigations, with particular emphasis on field imperfections and transmission characteristics \cite{bugrov2023modelling,douglas1999spatial,douglas2002influence,sysoev2022balance,bracco2008comparison,douglas2009linear}. In practical implementations, hyperbolic electrodes are commonly replaced by circular rods to facilitate machining. However, this substitution introduces higher-order multipole components, most dominantly the dodecapole and icosapole terms which degrade the ideal quadrupole performance\cite{douglas2002influence,sysoev2022balance,JANA2025117495}. Previous studies have demonstrated that selecting a rod-to-field radius ratio ($\eta$) close to $1.14$ minimizes the dodecapole contribution and improves mass resolution \cite{dayton1954measurement,reuben1996ion,douglas1999spatial,gibson2001numerical}.

\begin{figure}
    \centering
    \includegraphics[width=1\linewidth]{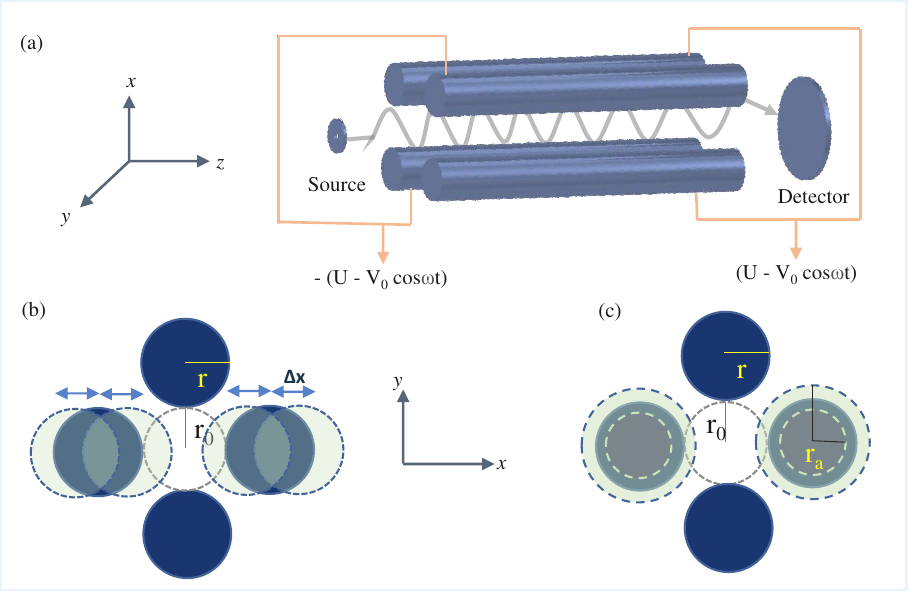}
    \caption{(a) Schematic of the linear QMF with circular rods and electrical connections. Sectional view ($xy$ plane) of the QMF with $x$-electrodes (b) displaced by $\Delta x$ from their normal position, (c) chosen of radius different from other set ($r-r_a=\Delta x$).}
    \vspace{-0.5cm}
    \label{fig:schematic}
\end{figure}

Deviations from this ideal QMF geometry, whether from machining tolerances or intentional design choices, introduce additional multipole field components such as the octupole and hexadecapole, thereby altering transmission characteristics \cite{konenkov2007mass,sudakov2003linear,taylor2008prediction}. For example, Ding et al. demonstrated that introducing a controlled $2.6\%$ octupole component—achieved by fabricating one pair of rods with a slightly larger diameter than the other pair and connecting the smaller rods to the positive DC output—led to a significant improvement in resolution~\cite{ding2003quadrupole}. More recently, Sysoev et al. reported that a small displacement of a diametrically opposite pair of electrodes effectively balances the amplitudes of the sixth and tenth-order harmonics, resulting in enhanced resolution~\cite{sysoev2022balance}. While certain controlled distortions can be beneficial, it has also been shown that an imperfection in a single electrode degrades resolution~\cite{JANA2025117495}.

In this work, we systematically investigate the influence of radial asymmetry in a linear QMF, shown schematically in Figure~\ref{fig:schematic}(a), by considering two distinct perturbations: displacement of a diagonal electrode pair (Figure~\ref{fig:schematic}(b)) and variation in electrode radii (Figure~\ref{fig:schematic}(c)). A potential model is developed to describe the resulting field distortions, from which the first stability region is extracted and analyzed. Building on this framework, we systematically examine the transmission characteristics of asymmetric configurations to elucidate how the asymmetry parameter modifies the transmission peak position and resolution. Particular attention is given to the role of higher-order multipole contributions in shaping performance, including the identification of conditions under which controlled asymmetry can be leveraged to enhance resolution.

\section{Theoretical Framework}
The potential within the QMF is altered when asymmetry is introduced between the transverse directions in the radial plane, leading to corresponding modifications in the stability region. In this section, we develop a potential model for a radially asymmetric QMF and extract the first stability region of ion motion.

The most general form of the potential at any point ($x,y$) in a quadrupolar device can be expressed with reference to rectangular coordinates as 
\begin{equation}\label{eq1}
    \phi(x,y)=\alpha(x^2-y^2)+C,
\end{equation}
where $\alpha$ and $C$ both are independent of $x$ and $y$. While $\alpha$ includes the RF potential either alone or in addition to the DC potential applied between the electrodes of opposing polarity, $C$ includes the same but applied effectively to all the electrodes~\cite{march2005quadrupole}. The potential on the surface of the $x$-pair of electrodes ($x=x_0$, $y=0$) and $y$-pair of electrodes ($x=0$, $y=y_0$) are respectively given by
\begin{equation}\label{eq2}
    \phi_x=\alpha x_0^2+C, \; \phi_y=-\alpha y_0^2+C
\end{equation}
The difference between these two potentials, the actual quadrupolar potential to which an ion is subjected, is thus 
\begin{equation}\label{eq3}
    \phi_0=\phi_x-\phi_y=\alpha(x_0^2+y_0^2). 
\end{equation}
Hence, by eq.~\ref{eq1}
\begin{equation}\label{eq4}
    \phi(x,y)=\phi_0\frac{(x^2-y^2)}{(x_0^2+y_0^2)}+C.
\end{equation}
For $\phi_x=-\phi_y=U-V_0\cos\omega t$, $\phi_0=\phi_x-\phi_y=2(U-V_0\cos\omega t)$ and hence the potential at any point inside the QMF at an instant can be defined from eq.~\ref{eq4} as
\begin{equation}\label{eq5}
    \phi(x,y,t)=2\frac{(x^2-y^2)}{(x_0^2+y_0^2)}(U-V_0\cos\omega t)+C.
\end{equation}
The equations of motion of an ion of charge $e$ and mass $m$ in the radial plane under the influence of this potential are described by the Mathieu differential equation as
\begin{equation}\label{eq6}
  \frac{d^2 u}{d\tau^2} + (a_u - 2q_u \cos 2\tau)u = 0,
\end{equation}
where \(u=x,y\) and 
\begin{equation}\label{eq7}
    a_x =\frac{8eU}{m x_0^2 \omega^2}, \quad a_y =-\frac{8eU}{m y_0^2 \omega^2}, \quad q_x = \frac{4eV_0}{m x_0^2 \omega^2}, \quad q_y = -\frac{4eV_0}{m x_0^2 \omega^2}, \quad \tau=\omega t/2.
     \end{equation}
In the symmetric configuration $x_0=y_0=r_0$, the radius of the inscribed circle while in the asymmetric setup, $y_0=r_0$ and $x_0=r_0+\Delta x$ as shown schematically in Figure~\ref{fig:schematic}(b), (c). The equations of motion inside the asymmetric quadrupole can be expressed in normalized form by equation \ref{eq6} as 
\begin{equation}\label{eq8}
\ddot{x} + \frac{a - 2q \cos2\tau}{p}x = 0, \quad \ddot{y} - \frac{a - 2q \cos2\tau}{p}y = 0,
\end{equation}
where
\begin{equation*}
a=\frac{8eU}{m r_0^2 \omega^2}, \quad q=\frac{4eV_0}{m r_0^2 \omega^2}; \quad \mathtt{and} \quad p = \frac{1 + \left(1 + \frac{\Delta x}{r_0}\right)^2}{2}.
\end{equation*}
is the geometric correction factor accounting for the radial asymmetry (see supplementary section). Numerical integration of these equations was performed using the Runge--Kutta 5(4) method (RK45) \cite{DORMAND198019,yang2015runge,tadmor2025stability} over the interval $\tau \in [0, 200]$ with 10,000 equally spaced time steps. The initial conditions were $(x, \dot{x}, y, \dot{y}) = (0, 1, 0, 1)$ \cite{funada2005solution}, and the system was considered stable if $|x|, |\dot{x}|, |y|, |\dot{y}| < 200$ throughout the integration. The parameter space was explored for $a \in [-2, 2]$ and $q \in [0, 2]$ with 1000 equally spaced values in each range, and calculations were repeated for $r_0 = 5.0$~mm with $\Delta x=0,\pm 0.02_0, \pm 0.04r_0, \pm 0.06r_0$ computing $p$ in each case.

\begin{figure}
    \centering
    \includegraphics[width=0.55\linewidth]{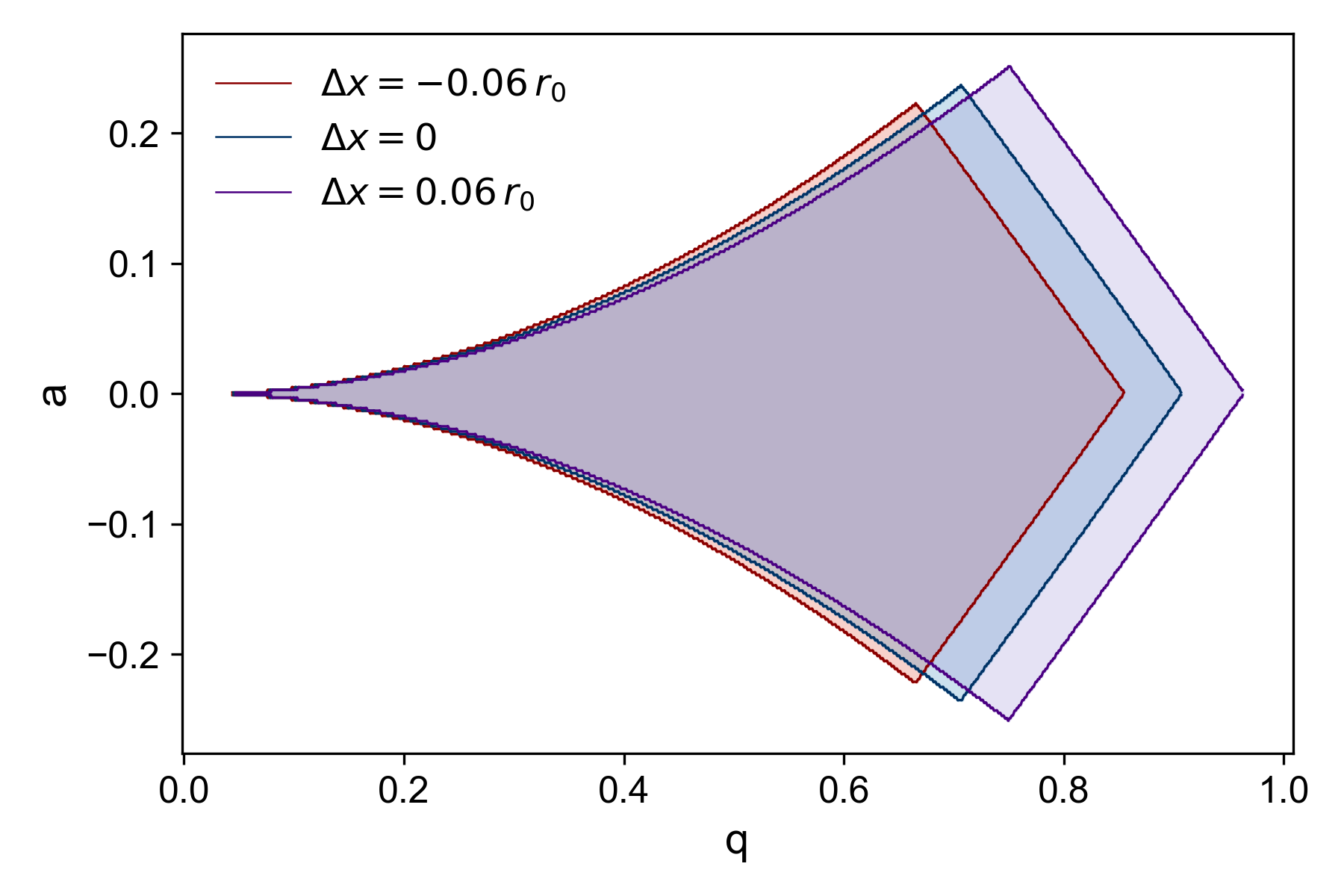}
    \caption{First stability region of radially asymmetric QMFs for $\Delta x=\pm0.06r_0$ in comparison to a symmetric QMF ($\Delta x=0$) as obtained from RK45 integration of the normalized equations of motion (Eq.~\ref{eq8}) in the \((a,q)\) plane.}
    \label{fig:stability}
\end{figure}

Figure~\ref{fig:stability} presents the first stability region for the symmetric configuration ($\Delta x = 0$) alongside asymmetric configurations with $\Delta x=\pm0.06r_0$. A distinct shift in the apex position of the stability boundary ($q_0$) is observed: contraction of the $x$ radius ($\Delta x<0$) causes $q_0$ to shift toward lower $q$ values, whereas expansion of the $x$ radius ($\Delta x>0$) results in a shift toward higher $q$ values. In addition, area of the stability region increases systematically with an increase in the effective inner radius of the QMF.

\section{Results and Discussion} 

Radial asymmetry in the QMF can be introduced by displacing, or by varying the radius of one diagonal pair of electrodes, as schematically illustrated in figure~\ref{fig:schematic}(b), (c). In order to quantify the degree of asymmetry, we define the parameter $\gamma_d=\Delta x/r_0$ corresponding to the displacement $\Delta x$ of each electrode along the $x$-axis from its normal position in the symmetric configuration~(figure \ref{fig:schematic}(b)). Similarly, another asymmetry parameter is defined as $\gamma_r=(r-r_a)/r_0$ corresponding to the variation in the radius of the electrode pair along the $x$-axis ($r_a$) as compared to the radius of the electrodes ($r$) in the symmetric design~(figure \ref{fig:schematic}(c)). For a given $r_0$, identical values of $\gamma_d$ and $\gamma_r$ correspond to the same radial asymmetry $\Delta x$ along the $x$-direction. However, these two modes of asymmetry perturb the multipole fields differently and, consequently, lead to distinct effects on the transmission characteristics of the QMF.

\subsection{Multipole field analysis}
QMFs employing circular rods inherently exhibit higher-order field components, with the dodecapole and icosapole terms being the most prominent beyond the fundamental quadrupole contribution\cite{douglas2002influence,sysoev2022balance}. The introduction of geometric asymmetry in the electrode configuration leads to the emergence of additional multipole terms, notably the octupole and hexadecapole components, as reported in previous studies on QMF\cite{JANA2025117495,douglas2009linear,ding2003quadrupole,sudakov2003linear} and experimental investigation with linear ion trap~\cite{mandal2024non}. In continuation of the preceding section, the general form of the multipole potential in the radial plane of the asymmetric QMF is given by  

\begin{equation}\label{eq9}
    \phi(x,y,t)=\sum_{N=1}^{\infty}\phi_{2N}(x,y)(U-V_0\cos\omega t),
\end{equation}
where 
\begin{equation}\label{eq10}
    \phi_{2N}(x,y)=\mathtt{Re}\sum_{N=1}^{\infty}\ A_{2N}\left(\frac{2z^2}{x_0^2+y_0^2}\right)^{N}, \,z=x+iy.
\end{equation}
Here, we have ignored $C$ as it does not appear in the equation of motion. The quadrupole, octupole, dodecapole, hexadecapole and icosapole terms correspond to $N=1,2,3,4$ and $5$ respectively.

The influence of geometric asymmetry in the QMF design on higher-order multipole fields is investigated using SIMION\cite{dahl2000simion,brabeck2016computational,hu2024simulation} simulations, following the methodology established in our previous work\cite{JANA2025117495}. The potential surface inside the QMF is simulated by applying a DC potential of $+1$~V to one diagonal pair of rods and $-1$~V to the opposite pair. For all electrode configurations considered in this study, $r_0$ was fixed $5.0$~mm. The nominal electrode radius, $r$ was calculated according to $r=\eta r_0$. The displacement of a rod from its symmetric position was defined as $\delta x=\gamma_d r_0$. For asymmetric designs, the radius of the modified electrode $r_a$ was evaluated using the relation $r_a=r-\gamma_r r_0$. The simulated potential is fitted to eq.~\ref{eq10} for $N$ values from $1$ to $5$, and thereby the coefficients $A_{2N}$ have been determined. 

\begin{table*}[h!]
\centering \label{table1}
\caption{\label{asym-coeff}Amplitudes of the multipole terms $A_{2N}$ for $\eta=1.12$, $\gamma_d=\gamma_r=0.02$}
\begin{tabular}{c|cccccc}
\hline
Asymmetry & $A_2$ & $A_4 \times 10^3$ & $A_6 \times 10^3$ & $A_8 \times 10^3$ & $A_{10} \times 10^3$ \\  
\hline
$\gamma_d=0.02$, ref. \cite{sysoev2022balance} & 0.9805  & 0.49   & 2.25   & 0.426  & -2.19   \\
$\gamma_d=0.02$, this work    & 1.00    & 0.613  & 2.37   & 0.481  & -2.42   \\
$\gamma_r=0.02$, this work    & 0.999   & 2.33   & 3.05   & 0.556  & -2.40  \\
\hline
\end{tabular}
\label{table1}
\end{table*}

Table~1 presents the multipole coefficients determined in our work for $\eta=1.12$ and $\gamma_d=\gamma_r=0.02$, and compares with the values previously reported by Sysoev et al.\cite{sysoev2022balance} for $\eta=1.12$ and $\gamma_d=0.02$. The $A_{2N}$ values obtained in our work for $\gamma_d=0.02$ are in good agreement with those reported previously \cite{sysoev2022balance}. The small deviations can be attributed to the precision of the potential surface simulations and the potential formulation adopted in our study. While $A_6$, $A_8$ and $A_{10}$ remain comparable, it is noteworthy that the octupole coefficient $A_4$ increases significantly when radial asymmetry is introduced by varying the electrode radii, as opposed to displacing the diagonal electrode pair ($\gamma_d=\gamma_r=0.02$, Table 1). 

\begin{figure}[h!]
    \centering
    \begin{minipage}[t]{0.49\textwidth}
        \begin{overpic}[width=\textwidth]{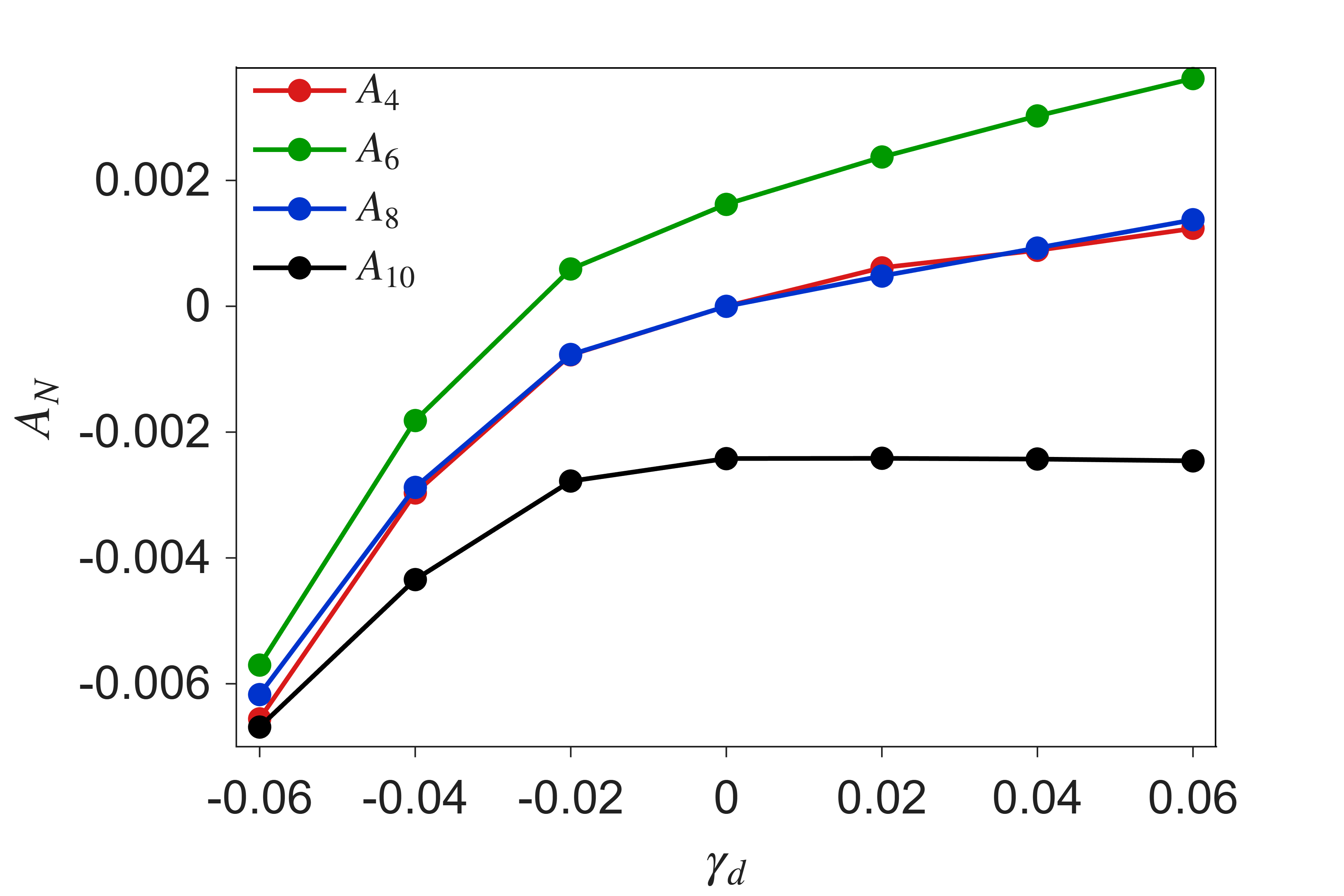}
            \put(1,60){\small\textbf{(a)}} 
        \end{overpic}
    \end{minipage}
    \hspace{-6mm}%
    \begin{minipage}[t]{0.49\textwidth}
        \begin{overpic}[width=\textwidth]{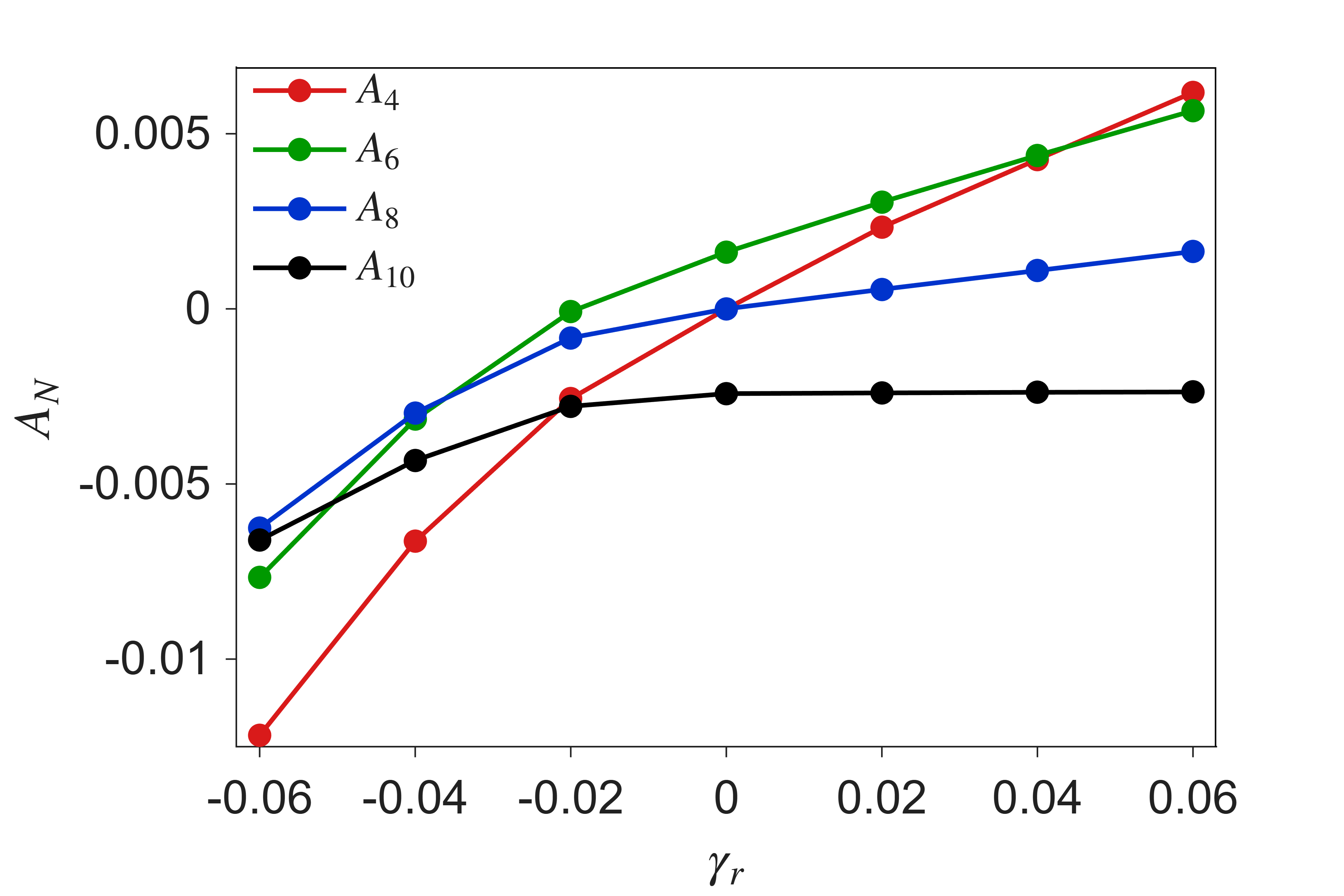}
            \put(1,60){\small\textbf{(b)}} 
        \end{overpic}
    \end{minipage}
    \caption{Variation of the multipole coefficients with (a) $\gamma_d$ and (b) $\gamma_r$ at $\eta=1.12$. The solid line serves as a guide to the eye only.}
    \label{fig:An}
\end{figure}

A systematic investigation was performed by varying both $\gamma_d$ and $\gamma_r$ within the range $-0.06\leq\gamma_{d,r}\leq 0.06$ for $\eta=1.10, 1.12$ and $1.14$, in order to examine the dependence of the multipole coefficients on the radial asymmetry parameters. Figures~\ref{fig:An}(a) and \ref{fig:An}(b) illustrate the variation of $A_{2N}$ as a function of $\gamma_d$ and $\gamma_r$ respectively, for $\eta=1.12$. As observed in figure~\ref{fig:An}(a), $|A_4|$, $|A_6|$ and $|A_8|$ increase monotonically when a diagonal pair of electrodes is displaced either inward or outward from their nominal position in the symmetric configuration. In contrast, $|A_{10}|$ increases for the inward displacement (i.e. $\gamma_d<0$) but remains nearly unchanged for outward displacement (i.e. $\gamma_d>0$). A similar qualitative trend is obtained when the radii of the diagonal electrode pair are varied. However, $|A_4|$ exhibits a steeper increase with $\gamma_r$ as compared to $\gamma_d$, consistent with an earlier study on asymmetry introduced by radius variation~\cite{sudakov2003linear}. 

The $A_{2N}$ coefficients exhibit similar trends for $\eta=1.14$ and $1.10$. It is worth noting, while identical values of $\gamma_r$ and $\gamma_d$ correspond to the same degree of radial asymmetry in the mass filter geometry for a given $\eta$, they distort the quadrupolar field in different ways, leading to distinct effects on ion transmission, as discussed in the following section.

\subsection{Transmission characteristics}

In order to investigate the effect of radial asymmetry on the transmission characteristics of QMFs, a series of simulations were performed for various geometries by systematically varying the parameters $\gamma_d$ and $\gamma_r$ at $\eta=1.10, 1.12$ and $1.14$ using SIMION. In each simulation, a beam of 1000 ions with a mass-to-charge ratio $40$~u/C, uniformly distributed in time of birth between $0$ and $0.5$~$\mu$s and spatially distributed over a circular cross-section of radius $0.1r_0$ was transmitted through the QMF of length $80$~mm, operated at a frequency $2$~MHz. This configuration ensured that the ions entered the QMF at all possible phases of the RF field and experienced more than $100$ RF cycles during transmission, consistent with the approach adopted in our previous work\cite{JANA2025117495}. The scan parameter, defined as $\lambda=a/2q$ was fixed at $0.16658$ for all transmission contour simulations.

\begin{figure}[t!]
    \centering

    \begin{minipage}[t]{0.495\textwidth}
    \begin{overpic}[width=\textwidth]{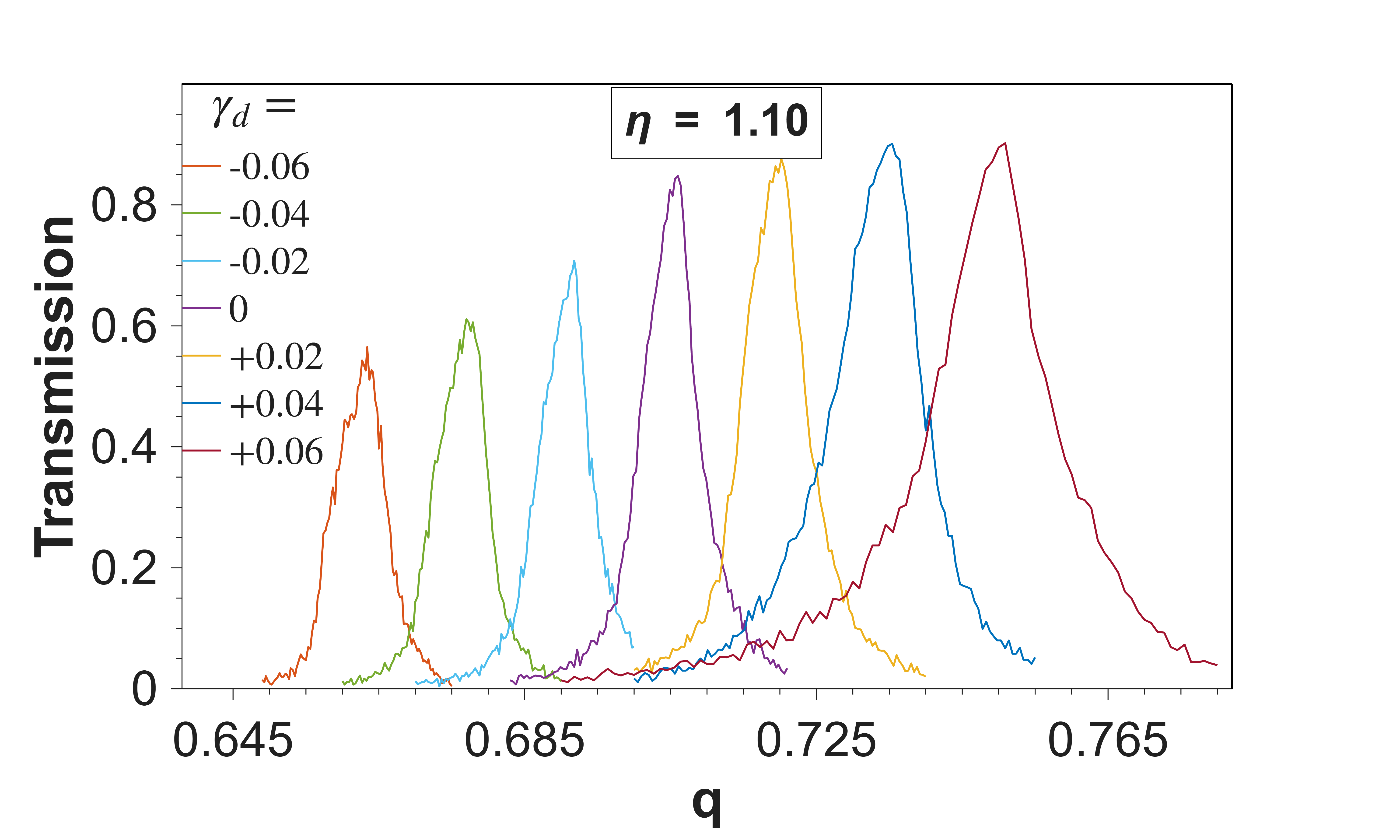}

        \put(3,50){\small\textbf{(a)}}
    \end{overpic}
\end{minipage}\hspace{-8mm}%
\begin{minipage}[t]{0.495\textwidth}
    \begin{overpic}[width=\textwidth]{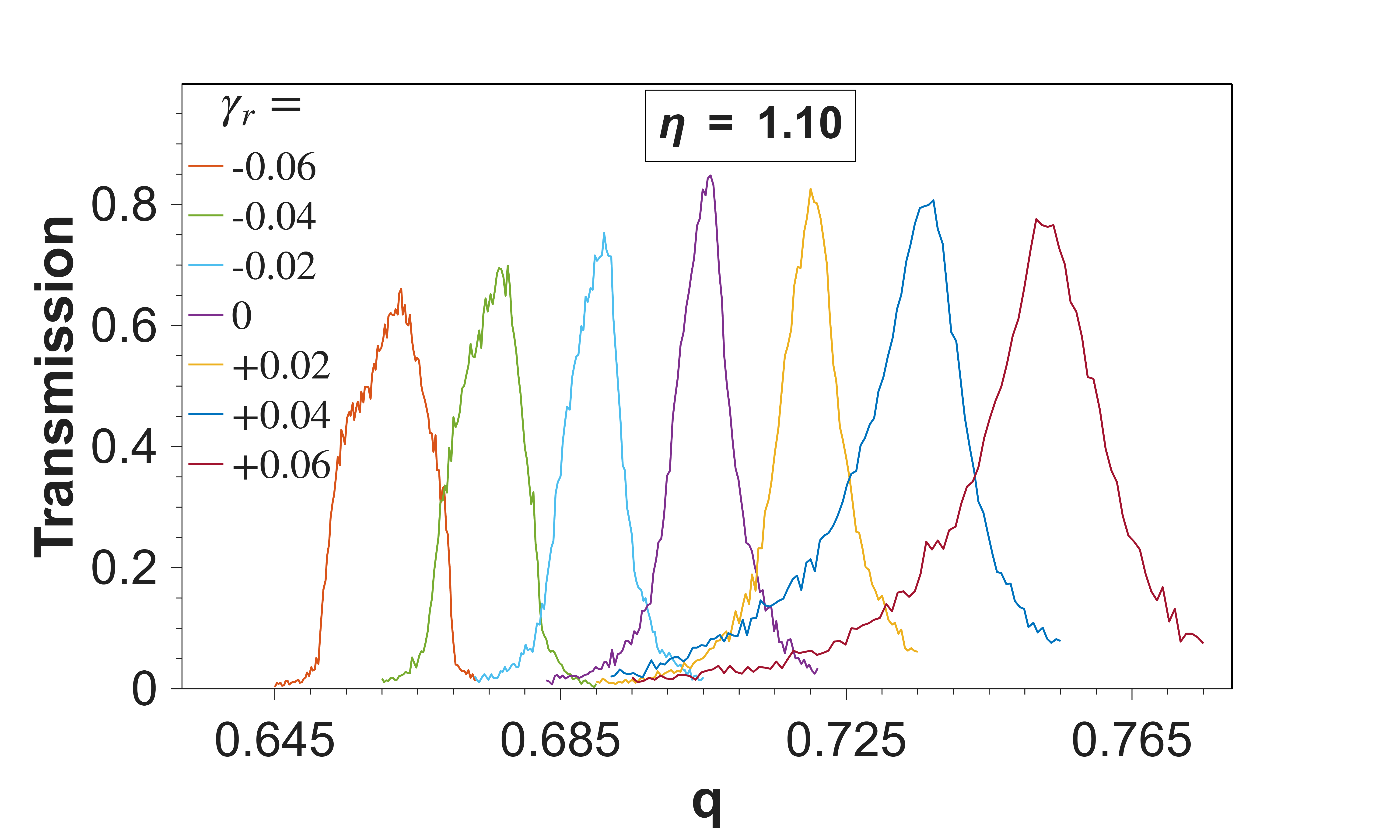}
        \put(3,50){\small\textbf{(b)}}
    \end{overpic}
\end{minipage}%
\vspace{-1.1mm}
    \begin{minipage}[t]{0.495\textwidth}
        \begin{overpic}[width=\textwidth]{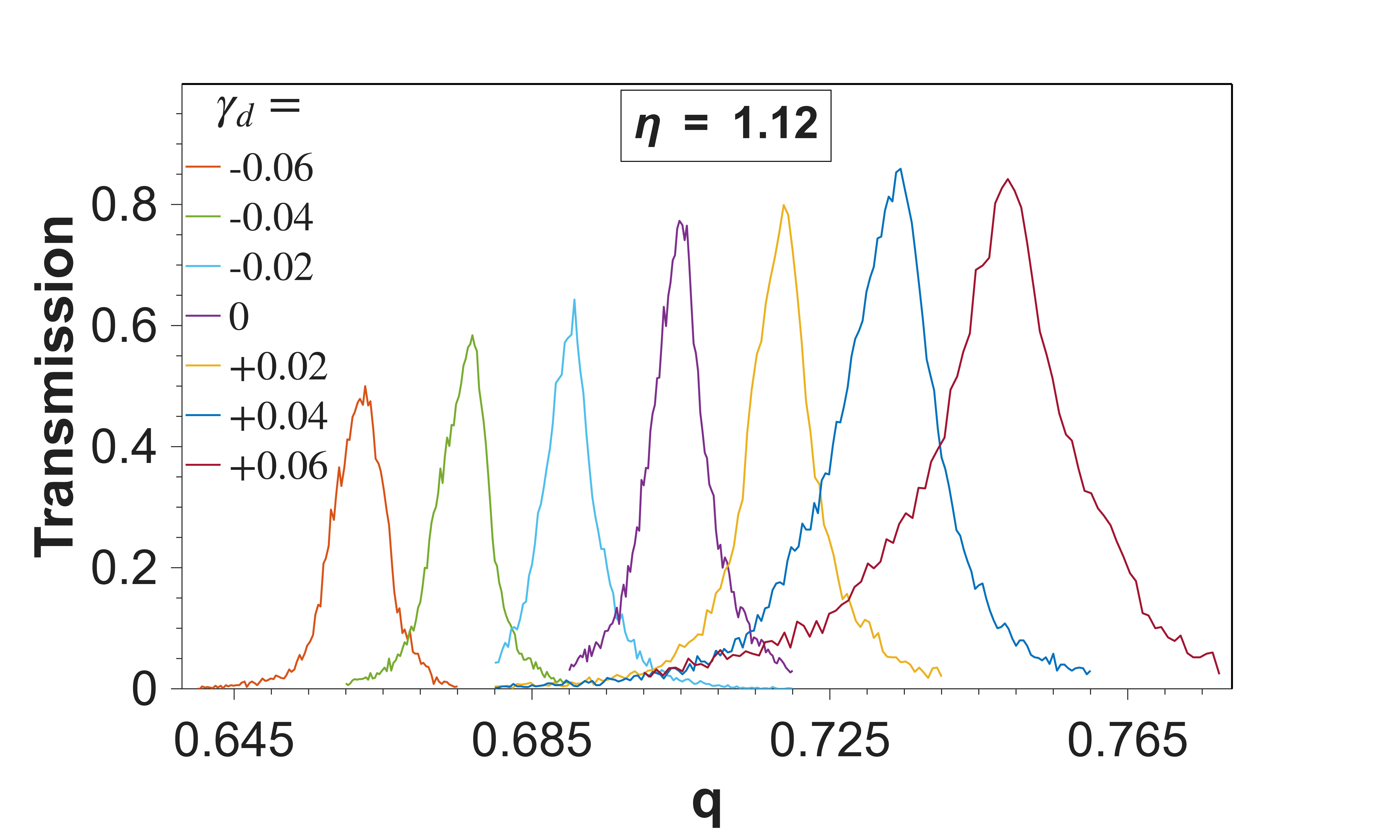}
            \put(3,50){\small\textbf{(c)}}
        \end{overpic}
    \end{minipage}\hspace{-8mm}%
    \begin{minipage}[t]{0.495\textwidth}
        \begin{overpic}[width=\textwidth]{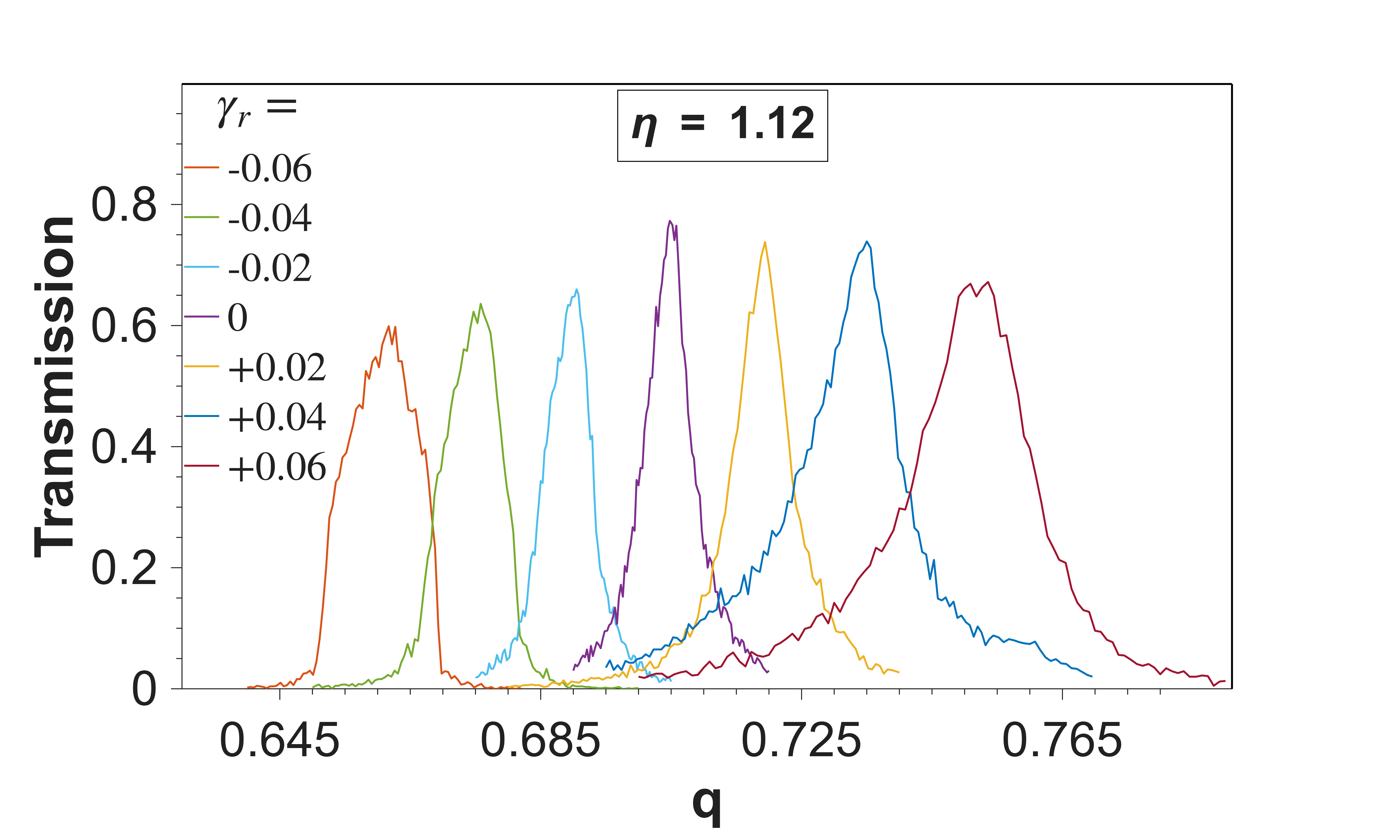}
            \put(3,50){\small\textbf{(d)}}
        \end{overpic}
\end{minipage}
\vspace{-3mm}
    \begin{minipage}[t]{0.495\textwidth}
        \begin{overpic}[width=\textwidth]{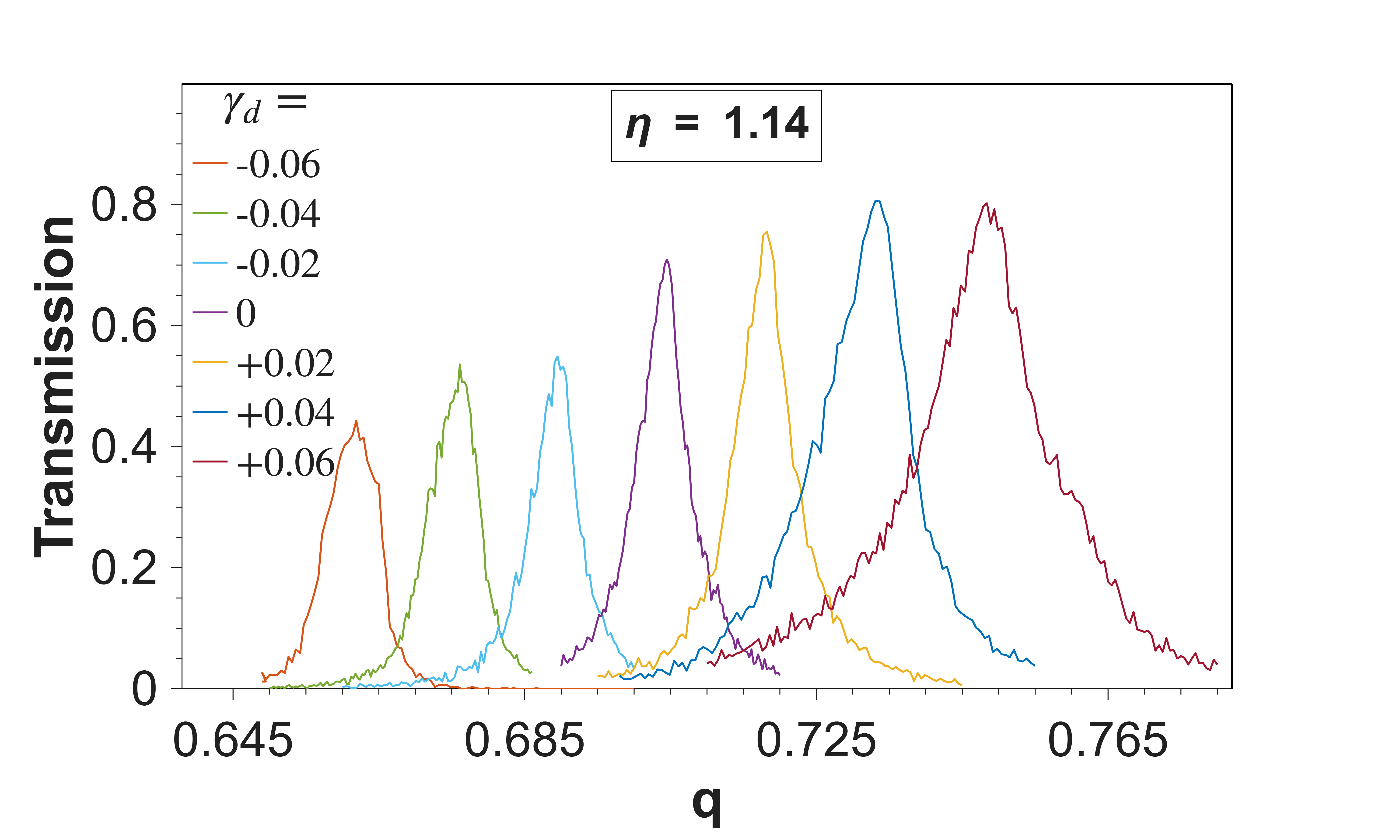}
            \put(3,50){\small\textbf{(e)}}
        \end{overpic}
    \end{minipage}\hspace{-8mm}%
    \begin{minipage}[t]{0.495\textwidth}
        \begin{overpic}[width=\textwidth]{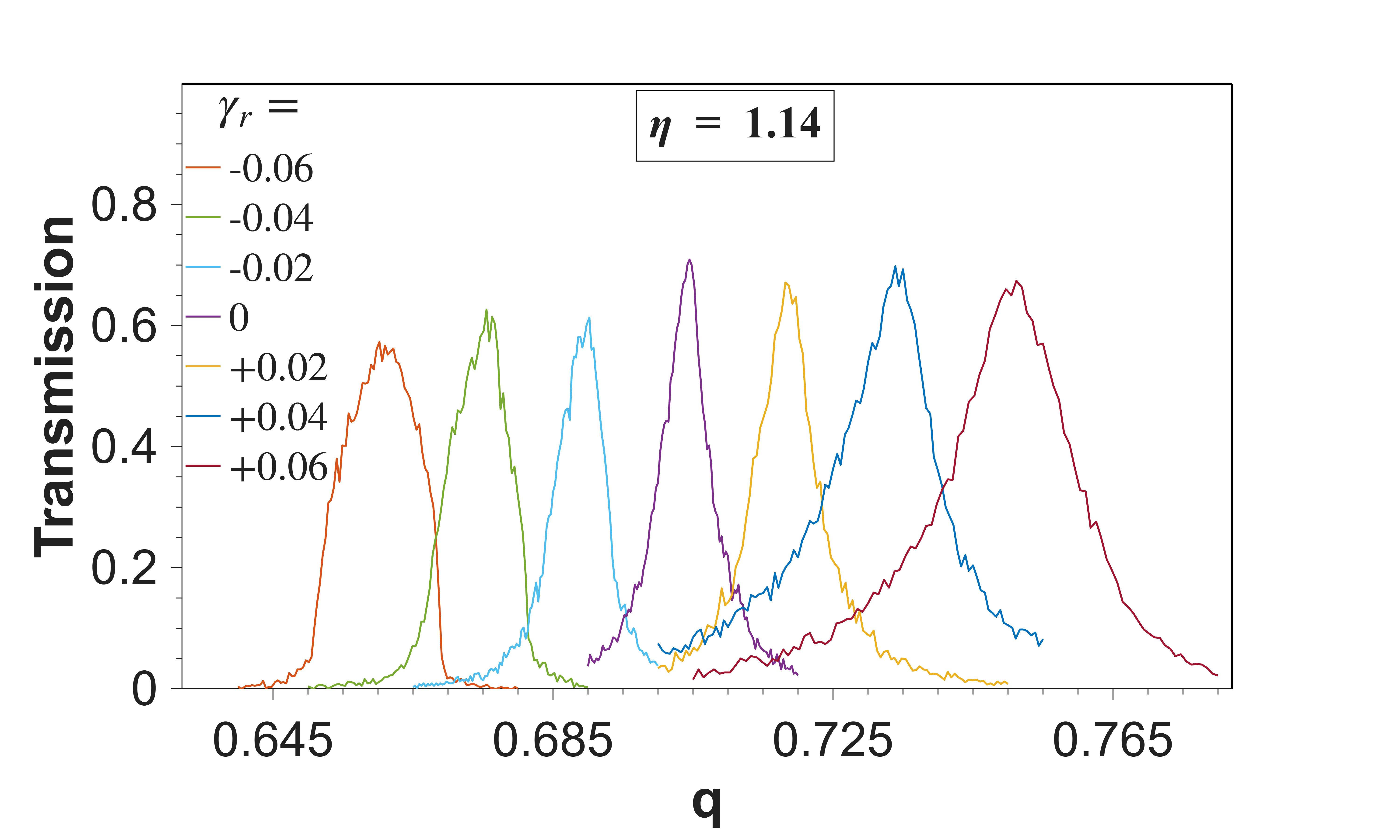}
            \put(3,50){\small\textbf{(f)}}
        \end{overpic}
    \end{minipage}

    \caption{Transmission contours for various values of $\eta$.  
    Displacement asymmetry results are shown in (a), (c), and (e),  
    while the radial asymmetry plots are shown in (b), (d), and (f), for $\eta=1.10, 1.12$ and $1.14$ respectively. Each transmission contour was generated using 100 evenly spaced data points.}
    \label{fig:transmission_plots}
\end{figure}

Figures~\ref{fig:transmission_plots}(a), (c) and (e) show the transmission contours through the QMF for electrode displacements in the range $-0.06r_0\leq \Delta x\leq0.06r_0$ at $\eta=1.10,1.12$, and $1.14$, respectively. In all three cases, a consistent trend is observed: inward displacement of the electrodes results in a shift of the transmission peak toward lower $q$ values accompanied by a reduction in transmission efficiency, whereas outward displacement causes the peak to shift toward higher $q$ values with a corresponding increase in efficiency. Similarly, figures \ref{fig:transmission_plots}(b), (d), and (f) show the transmission contours obtained by varying the radius of one diagonal pair of electrodes within $-0.06r_0$ to $0.06r_0$ at the same $\eta$ values. The transmission peak shifts exhibit trends consistent with those observed for electrode displacement.

\newcommand{\figpanel}[3]{
    \begin{minipage}[t]{#3\textwidth}
        \raggedright\textbf{#1})\\[-1pt] 
        \includegraphics[width=\linewidth]{#2}
    \end{minipage}%
}

\begin{figure}[h!]
    \centering
    \includegraphics[width=0.5\linewidth]{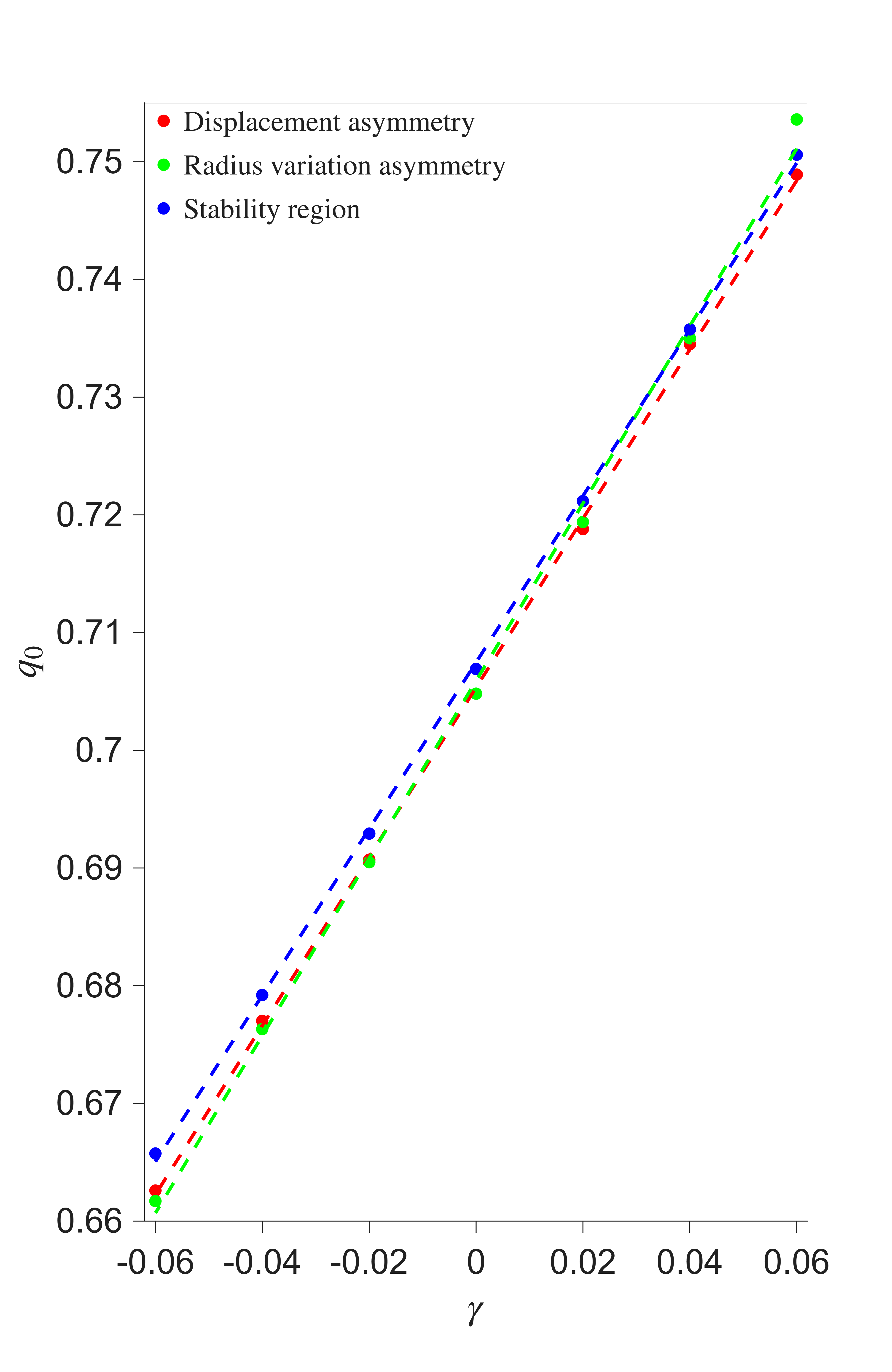}
    \caption{Variation of the apex of the stability diagram and the transmission peak ($q_0$) as a function of asymmetric parameter, $\gamma$, for $\eta$ = 1.12. The dashed lines show the linear fit in each case, with $R$-square value of 0.99968, 0.99806 and 0.99971 corresponding to displacement asymmetry, radius variation asymmetry and stability region respectively.}
    \label{fig:q0}
\end{figure}

A detailed analysis reveals that the transmission peak ($q_0$) exhibits a linear shift with the asymmetry parameters $\gamma_d$ and $\gamma_r$, as shown in figure~\ref{fig:q0} for $\eta = 1.12$. The apex of the first stability region, determined from the quadrupole potential model developed in this work for a radially asymmetric QMF, also displays a linear shift in $q$-space with increasing asymmetry (see figure \ref{fig:q0}). A strong correspondence is observed between the transmission peak shift caused by electrode displacement and the apex shift of the stability boundary, with a standard deviation of $0.0018$ for $\eta = 1.12$. An equally strong correlation is obtained for electrode radius variation, with a standard deviation of $0.0026$ for the same $\eta$ value. Similar correlations are found for other values of $\eta$ (e.g., $1.10$ and $1.14$) across both types of radial asymmetry. These results demonstrate that the observed shift of the transmission contour primarily originates from modifications to the underlying quadrupolar potential in a radially asymmetric QMF, rather than from the influence of higher-order multipole contributions.

\begin{figure}
    \centering
    \includegraphics[width=0.7\linewidth]{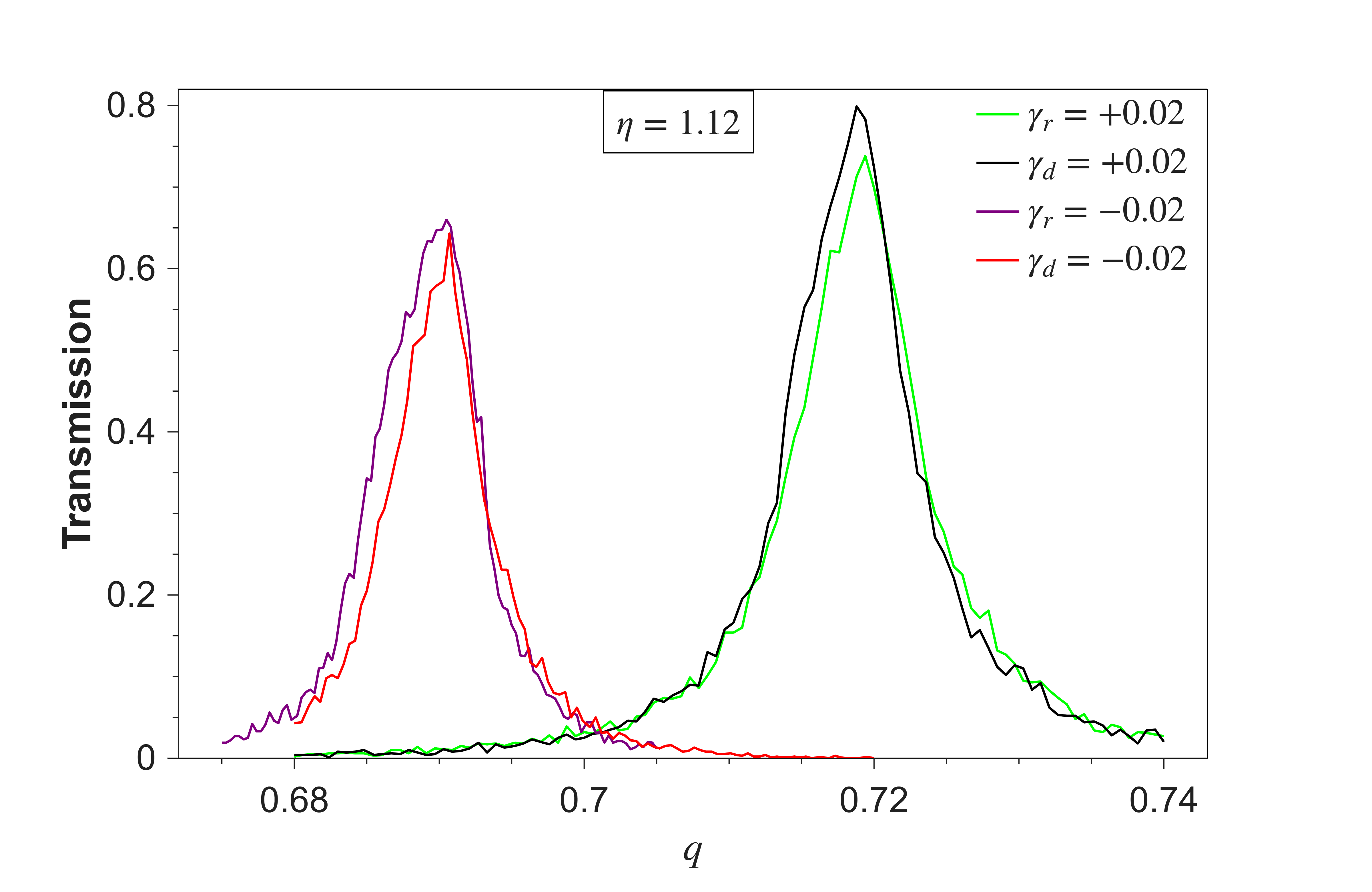}    
    \caption{Comparison of transmission profiles for displacement asymmetry ($\gamma_d = \pm 0.02$) and radius variation asymmetry ($\gamma_r = \pm 0.02$) at $\eta = 1.12$. Each transmission contour is based on 100 uniformly spaced data points.}
    \label{width}
\end{figure}

Although the transmission peaks for displacement asymmetry and radius variation asymmetry nearly coincide for $|\gamma|\leq0.04$ (figure~\ref{fig:q0}), the corresponding transmission contours display a noticeable difference, as seen in figure~\ref{width} for $\gamma_d=\gamma_r=\pm 0.02$ at $\eta=1.12$. This difference arises mainly from the variation in the strengths of the multipole fields generated in the two cases for the same asymmetry parameter ($\gamma_d=\gamma_r$), which in turn influences the resolution differently, as discussed in the following section.

\subsection{Resolution}

\begin{figure}
    \centering
    \begin{overpic}[width=0.65\textwidth]{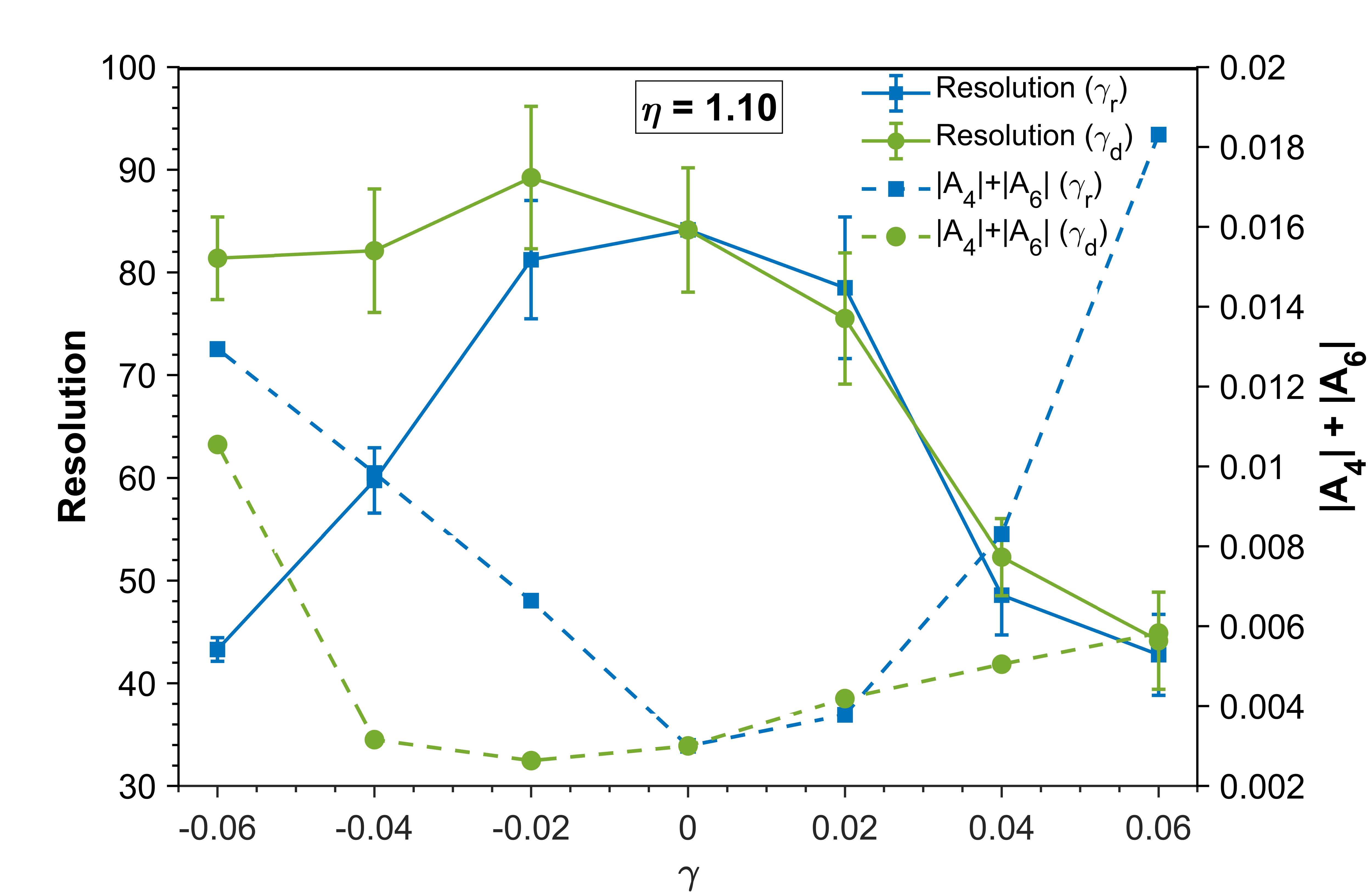}
        \put(2,58){\small\textbf{(a)}}
    \end{overpic}
    \vspace{-0.2mm}%
    \\
    \begin{overpic}[width=0.65\textwidth]{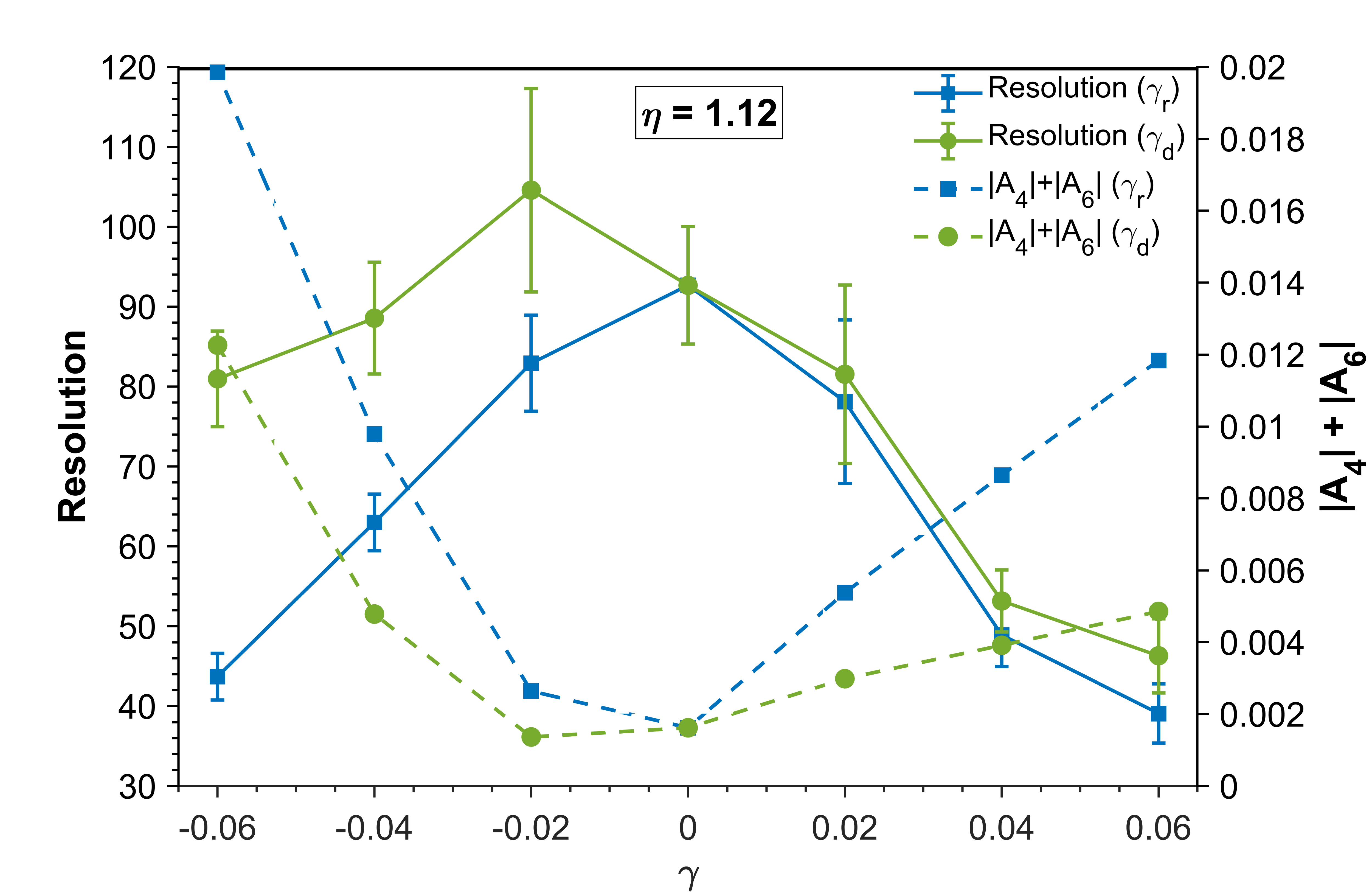}
        \put(2,58){\small\textbf{(b)}}
    \end{overpic}
    \vspace{-0.2mm}%
    \\
    \begin{overpic}[width=0.65\textwidth]{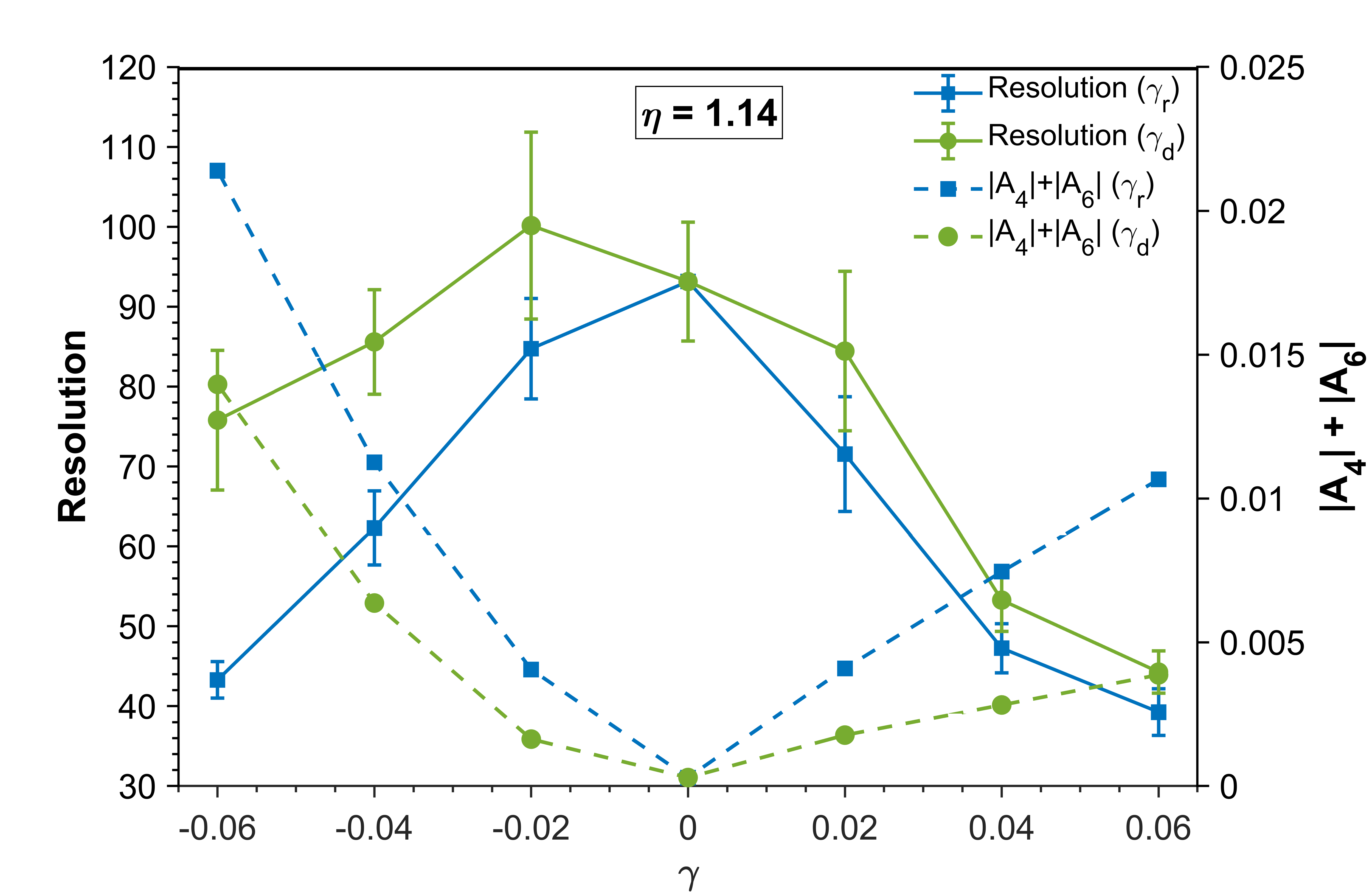}
        \put(2,58){\small\textbf{(c)}}
    \end{overpic}
    \caption{Resolution (left axis) and $|A_4|+|A_6|$ (right axis) as a function of $\gamma_d$ and $\gamma_r$ at (a) $\eta=1.10$, (b) $\eta=1.12$ and (c) $\eta=1.14$. The solid and dashed lines are included merely as a visual guide.}
    \label{fig:R-A4-A6}
\end{figure}

The resolution, $R$, of a QMF is influenced by several factors, including the scan parameter $\lambda$, the number of RF cycles experienced by the ions, and the electrode geometry\cite{syed2012factors,dawson2013quadrupole,march2005quadrupole}. 
As a conservative definition for comparing the performance of QMF of various asymmetric designs, we follow $R=q_0/\Delta q$ as introduced by Brabeck et al.\cite{brabeck2016computational} and employed in our earlier work\cite{JANA2025117495} as well as in other studies\cite{hu2024simulation}. Here $q_0$ is the peak position and $\Delta q$ is the full-width at half-maxima of the transmission contour. 

Following this definition, $R$ is evaluated from the transmission contours shown in figure~\ref{fig:transmission_plots} for both displacement and radius variation asymmetries. The dependence of transmission resolution, $R$, on the asymmetry parameters $\gamma_d$ and $\gamma_r$ is presented in figure~\ref{fig:R-A4-A6}(a), (b), and (c) for $\eta = 1.10$, $1.12$, and $1.14$, respectively.

In the case of radial asymmetry, introduced by symmetrically varying the radii of two opposite rods, the maximum resolution is obtained for the symmetric configuration (\(\gamma_r = 0\)). The resolution $R$ decreases progressively with increasing $|\gamma_r|$. This trend is consistent with our previous findings~\cite{JANA2025117495} where a similar decrease in resolution was observed for an asymmetric configuration in which the radius of a single rod was varied. 

An intriguing behavior is observed under displacement asymmetry, where the resolution $R$ reaches its maximum at a slight displacement ($\gamma_d=-0.02$) rather than at the nominal symmetric configuration ($\gamma_d=0$). Beyond this optimum, $R$ decreases for both larger inward displacements ($\gamma_d=-0.04, -0.06$) and outward displacements from the nominal position. This trend remains consistent across all three values of $\eta$ investigated in this work (1.10, 1.12 and 1.14). Such behavior suggests that a small degree of controlled displacement asymmetry may partially compensate higher-order multipole distortions, thereby enhancing transmission resolution before excessive asymmetry degrades stability.

This highlights the nuanced influence of electrode displacement on mass filter performance. To elucidate this effect, it is essential to examine the interplay of the multipole fields, particularly their impact on the stability diagram across different values of $\gamma_d$ and $\eta$. However, in the presence of higher-order multipole components, the boundaries of the stability diagram are not well defined\cite{douglas2009linear,noshad2011numerical}. Sysoev et al. reported~\cite{sysoev2022balance} a similar observation at $\eta=1.12$ only and argued that there is a nearly perfect balance between the 6th and 10th spatial harmonics amplitudes that yields improved resolution. However, it does not take into account the 4th and 8th spatial harmonics that appear due to rod displacement.

The octupole and dodecapole components emerge as the most significant distortion fields in a radially asymmetric quadrupole, since the former scales inversely with $r_0^4$ and the later with $r_0^6$. Our detailed analysis of the multipole coefficients reveals an empirical correlation between the combined amplitudes of these dominant harmonics, $|A_4|+|A_6|$, and the transmission resolution, $R$. Figure \ref{fig:R-A4-A6} (a), (b) and (c) illustrate the variation of $|A_4|+|A_6|$, with the asymmetry parameters $\gamma_d$ and $\gamma_r$. Notably, $|A_4|+|A_6|$ attains a minima at $\gamma_d=-0.02$ for $\eta=1.10$ and $1.12$ coinciding with the observed maximum in resolution. A consistent correlation is also observed for the radial asymmetry introduced by varying the radii of diametrically opposite electrodes, across $\eta=1.10, 1.12$ and $1.14$. In this case, the minimum of $|A_4|+|A_6|$, corresponding to the lowest field distortion, occurs at $\gamma_r=0$, i.e., the symmetric configuration, where the transmission resolution is likewise maximized. A similar correlation between multipole distortion and resolution was reported earlier in our study of an asymmetric QMF with a single rod of modified radius\cite{JANA2025117495}.

A mild deviation from the observed correlation between $|A_4|+|A_6|$ and $R$ is noted for displacement asymmetry at $\eta=1.14$. In this case, the resolution peaks at $\gamma_d=-0.02$, while $|A_4|+|A_6|$ attains its minimum at the symmetric configuration ($\gamma_d=0$). This can be attributed to the near-vanishing of $A_6$ at $\eta=1.14$ and the absence of $A_4$ in symmetric round-rod geometries, resulting in $|A_4|+|A_6|$ approaching zero at symmetry. The enhanced resolution observed at the `magic' displacement parameter $\gamma_d=-0.02$ may thus arise from a more complex interplay of multipole fields beyond the dominant octupole and dodecapole contributions at $\eta=1.14$.

\section{Conclusion}

This study models and analyzes the potential within a quadrupole mass filter incorporating radial asymmetry, introduced either by displacing or varying the radii of a diametrically opposite pair of electrodes. For the quadrupole field, the first stability region - critical for standard mass filter operation - was extracted, revealing that its apex shifts from the position observed in a conventional symmetric configuration. The asymmetric geometry also introduces additional multipole components, most notably the octupole and hexadecapole terms, in addition to the dodecapole and icosapole terms present in symmetric round-rod designs, with their strengths increasing proportionally to the degree of asymmetry.

Transmission simulations across a range of asymmetry parameters and rod-to-field radius ratios demonstrate that the transmission peak position shifts linearly with asymmetry. In most cases, transmission resolution decreases with increasing asymmetry, except for a ‘magic’ asymmetry parameter of $-0.02$ in the electrode displacement configuration, which yields a resolution improvement. The shift in the transmission contour arises predominantly from modifications to the quadrupolar potential in the asymmetric geometry, whereas resolution degradation is governed primarily by the combined influence of higher-order multipole fields, particularly the octupole and dodecapole components.

Our comprehensive analysis of the stability behavior and transmission characteristics of radially asymmetric QMFs further indicates that, in certain parameter regimes, maximum resolution can be achieved through a carefully engineered balance between geometric symmetry and controlled asymmetry. This finding opens new opportunities for precision tuning of mass filter performance in practical applications.

\begin{acknowledgement}
ND thanks SERB/ANRF India (CRG/2023/001529) and BRNS India (58/14/21/2023 - BRNS12329) for funding. SJ acknowledges Indian Association for the Cultivation of Science India for research fellowship.
\end{acknowledgement}

\section{Conflicts of Interest}
The authors declare no conflicts of interest for this manuscript.

\begin{suppinfo}
Additional theoretical derivations, numerical methods, and algorithmic details employed in the stability analysis are provided in the Supplementary Section.
\end{suppinfo}

\providecommand{\latin}[1]{#1}
\makeatletter
\providecommand{\doi}
  {\begingroup\let\do\@makeother\dospecials
  \catcode`\{=1 \catcode`\}=2 \doi@aux}
\providecommand{\doi@aux}[1]{\endgroup\texttt{#1}}
\makeatother
\providecommand*\mcitethebibliography{\thebibliography}
\csname @ifundefined\endcsname{endmcitethebibliography}  {\let\endmcitethebibliography\endthebibliography}{}

\end{document}